# Regional spread of chikungunya fever in the Americas (Supplementary Material – main paper in Eurosurveillance)

Simon Cauchemez, Martine Ledrans, Chiara Poletto, Philippe Quenel, Henriette De Valk, Vittoria Colizza, Pierre-Yves Boelle

**Contents**

# 1 Serial Interval distribution

The serial interval distribution is the time distribution between symptoms in the index case and that of its secondary cases. To account for the extrinsic period of transmission, it was described as a mixture of distributions based on the between bites duration (*i.e.* the gonotrophic cycle) and mortality rate in the mosquito [1]. In summary, secondary cases infections take place at time given by multiples of the gonotrophic cycle, starting with the infecting bite in the mosquito, with a weight decreasing due to mosquito mortality. On the basis of viremia time profile, a human host was assumed to be infectious for 5 to 8 days, with infectivity starting 1 to 2 days before symptoms. The gonotrophic cycle duration is largely determined by temperature (Otero) and is about 4 days in Saint-Martin with temperatures in the range around 26°C at this time of the year. This is also in accordance with infections in the same households, where about successive cases in the same household occurred roughly at multiple of 4 days (Figure S1).

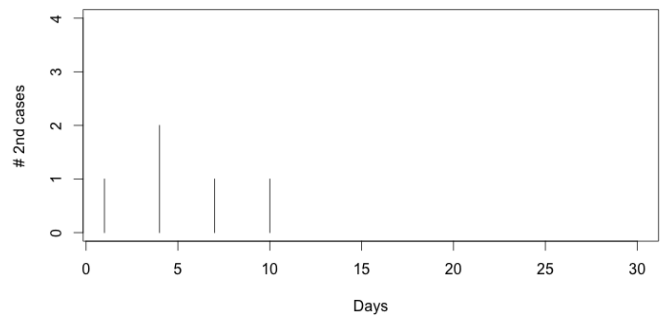

**Figure S1 : Observed serial intervals (symptoms to symptoms) in households in Saint-Martin. The first 1-day interval may be the result of simultaneous exposition.**

To account for the time it takes before the virus is found in the salivary glands of the mosquito, we considered that the first gonotrophic cycle after mosquito infection would not lead to transmission. The mortality rate of *A. aegypti* mosquitoes in the Caribbean was reported between 0.2 and 0.1 per day in Puerto Rico [2] – we selected 0.1/day. Finally, we only considered the first 10 gonotrophic cycles after infection, as few mosquitoes would survive more than 40 days (= 10 gonotrophic cycles). Overall, this led to a generation time distribution with mean 23 days and SD 6 days (Figure S2). We aggregated the serial interval distribution over week-long intervals, with the first interval only 4 days long so that the average duration was preserved (i.e. days 0-3, 4-11,…).

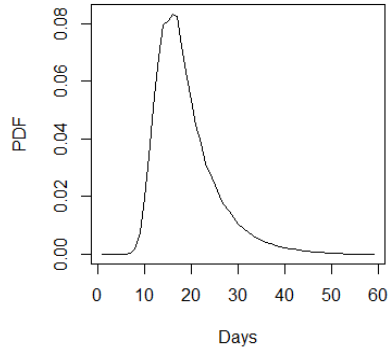

**Figure S2 : Probability density function of the serial interval distribution for chikungunya fever in the Caribbean.**

## 2   Reproduction number estimates in the three French islands

The reproduction ratio was estimated by the Exponential Growth method, i.e. $R = -1/\int_0^\infty \exp(-r\,\tau)\,g(\tau)d\tau$ where $g$ is the probability density function of the serial interval distribution and $r$ is the exponential growth rate (see Figure S2). To determine $R$, we fitted a Poisson regression to observed weekly incidence as a function of time and obtained $r$, then applied the transformation above. A large part of the variability in these estimates came from the time window selected for analysis, as exemplified in Figure S3. Indeed, the growth rate estimate changes and the R estimates are, from left to right, 1.6, 3.6 and 1.4.

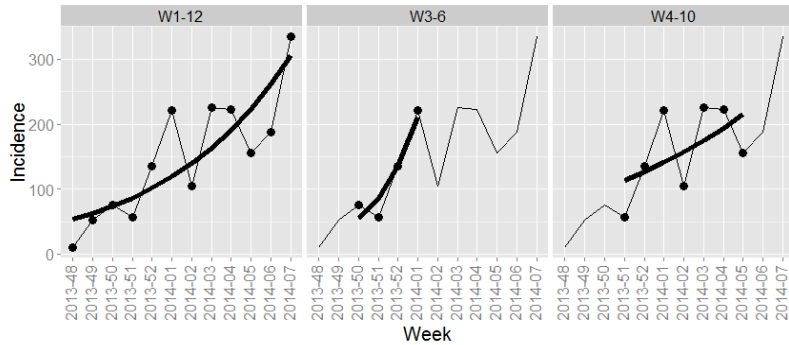

**Figure S3 : Three instances of exponential growth fitting to the epidemic curve in Saint-Martin, depending on the time window considered (from left to right: weeks 1-12, weeks 3-6, weeks 4-10)**

To overcome and illustrate this uncertainty, we reported the estimates obtained on the 10 periods where the exponential growth provided the best fit. The goodness of fit of the exponential growth regression was estimated by the deviance R-squared residual, computed as $R^2$ = (Dev(Null) – Dev(Exponential)) / Dev(Null) where Dev(Null) is the deviance of the null model computed as Dev(Null) = $2\sum_i y_i \ln\frac{y_i}{\bar{y}}$ and Dev(Exponential) is the deviance of the exponential model computed as Dev(Exponential) = = $2\sum_i (y_i - \hat{y}_i) + y_i \ln\frac{y_i}{\hat{y}_i}$ with $\hat{y}_i$ the fitted Poisson prediction. The results of this approach can be found in Table S1.

Sensitivity analyses were performed regarding the choice of the serial interval distribution. Using a shorter duration for the gonotrophic cycle (3 days vs. 4 days), led to little change in the serial interval distribution (2 days) and less than 5% variation on the estimates of $R$. With higher daily mortality in mosquitoes (15% instead of 10%), the generation interval was shorter, and the estimates of $R$ were reduced by ~20%.

**Table S1: Estimates of the reproduction number $R$ in the 3 French islands according to the choice of period for analysis.**

| Territory | $R$ | CI95% (lo,up) | | Deviance $R^2$ | begin | end |
|---|---|---|---|---|---|---|
| Guadeloupe | | | | | | |
| | 4,2 | 3,8 | 4,6 | 0,99 | 2014-06 | 2014-10 |
| | 4,5 | 4 | 5,1 | 0,99 | 2014-07 | 2014-10 |
| | 4,1 | 3,5 | 4,8 | 0,97 | 2014-06 | 2014-09 |
| | 3,4 | 3,1 | 3,6 | 0,96 | 2014-06 | 2014-11 |
| | 3 | 2,9 | 3,2 | 0,95 | 2014-05 | 2014-11 |
| | 3,3 | 3 | 3,6 | 0,94 | 2014-07 | 2014-11 |
| | 2,8 | 2,7 | 3 | 0,94 | 2014-04 | 2014-11 |
| | 3,4 | 3,1 | 3,7 | 0,93 | 2014-05 | 2014-10 |
| | 3 | 2,8 | 3,2 | 0,91 | 2014-04 | 2014-10 |
| | 3,1 | 2,8 | 3,5 | 0,9 | 2014-08 | 2014-11 |
| Martinique | | | | | | |
| | 2,5 | 2,3 | 2,6 | 0,99 | 2014-01 | 2014-06 |
| | 2,3 | 2,1 | 2,5 | 0,99 | 2014-01 | 2014-05 |
| | 2,7 | 2,4 | 3 | 0,99 | 2014-03 | 2014-06 |
| | 2,5 | 2,3 | 2,7 | 0,99 | 2014-02 | 2014-06 |
| | 2,3 | 2 | 2,7 | 0,98 | 2014-02 | 2014-05 |
| | 2,4 | 2 | 2,7 | 0,98 | 2014-01 | 2014-04 |
| | 2,2 | 2,1 | 2,3 | 0,96 | 2014-01 | 2014-07 |
| | 1,8 | 1,8 | 1,9 | 0,96 | 2014-01 | 2014-12 |
| | 1,7 | 1,7 | 1,8 | 0,96 | 2014-01 | 2014-14 |
| | 1,7 | 1,7 | 1,7 | 0,96 | 2014-02 | 2014-14 |
| Saint-Martin | | | | | | |
| | 4 | 3,4 | 4,8 | 0,89 | 2013-48 | 2014-01 |
| | 3,4 | 2,8 | 4,2 | 0,87 | 2013-49 | 2014-01 |
| | 4 | 3,1 | 5,1 | 0,86 | 2013-50 | 2014-01 |
| | 3,7 | 2,9 | 4,7 | 0,76 | 2013-48 | 2013-52 |
| | 1,6 | 1,5 | 1,6 | 0,72 | 2013-48 | 2014-09 |
| | 2,1 | 2 | 2,3 | 0,72 | 2013-48 | 2014-04 |
| | 1,5 | 1,5 | 1,6 | 0,72 | 2013-49 | 2014-09 |
| | 2,4 | 2,1 | 2,6 | 0,7 | 2013-48 | 2014-03 |
| | 1,9 | 1,7 | 2,1 | 0,68 | 2013-49 | 2014-04 |
| | 1,8 | 1,6 | 2,1 | 0,68 | 2014-05 | 2014-09 |

## 3  Model for a fully observed epidemic in the Caribbean

Data for the analysis of the spatial expansion of the chikungunya epidemic in the Caribbean are presented in Table S2.

Let's first assume that both the invasion status of territories and the times of invasion are known (we will show in the following section how to handle uncertainty about these variables).

We denote $T$ the time when the analysis is done. We denote $I_i = 1$ if territory $i$ is invaded; and $I_i = 0$ otherwise. The time of invasion of territory $i$ is denoted $t_i$. By convention, $t_i = T$ if a territory has not yet been infected at time $T$.

We denote $\lambda_{i \to j}$ the instantaneous risk of transmission from territory $i$ to territory $j$ as defined in Table 1.

The force of invasion exerted on territory $j$ at time $t$ is:

$$\Lambda_j(t) = \sum_{i:t_i<t} \lambda_{i \to j}$$

The contribution of territory $j$ to the likelihood is

$$P(I_j, t_j \mid \theta) = \begin{cases} \Lambda_j(t_j).\exp\left(-\int_0^{t_j} \Lambda_j(u)du\right) & \text{if } I_j = 1 \\ \exp\left(-\int_0^{t_j} \Lambda_j(u)du\right) & \text{otherwise} \end{cases}$$

The likelihood is:

$$P\left(\{I_j, t_j\}_j \mid \theta\right) = \prod_j \Lambda_j(t_j)^{I_j} .\exp\left(-\int_0^{t_j} \Lambda_j(u)du\right) \quad (1)$$

**Table S2 : Dates of publication of reports about the epidemiological situation in territories and source of information**

| Territory | Date | Source |
|---|---|---|
| **Territories with local transmission** | | |
| Saint-Martin | 9/12/13 | Local health authorities. http://www.invs.sante.fr/fr/content/download/81267/296431/version/43/file/pe_chikungunya_antilles_111213.pdf |
| Martinique | 19/12/13 | Local health authorities. http://www.invs.sante.fr/fr/content/download/81629/297952/version/44/file/pe_chikungunya_antilles_191213.pdf In a communication to promed (20140129.2240929) a suspect case was described as early as 2/11/13. http://www.promedmail.org/direct.php?id=2240929 |
| Guadeloupe | 28/12/13 | Local halth authorities. http://www.invs.sante.fr/fr/content/download/81802/298759/version/45/file/pe_chikungunya_antilles_261213.pdf |
| Saint-Barthelemy | 28/12/13 | Local health authorities. http://www.invs.sante.fr/fr/content/download/81802/298759/version/45/file/pe_chikungunya_antilles_261213.pdf |
| British Virgin Islands | 13/1/14 | Promed Mailing list (20140114.2172714) http://www.promedmail.org/direct.php?id=2172714 |
| Dominica | 16/1/14 | Promed Mailing List (20140118.2181292 , 20140129.2240929) http://www.promedmail.org/direct.php?id=2181292 http://www.promedmail.org/direct.php?id=2240929 |
| Anguilla | 7/2/14 | Promed Mailing list. (20140207.2262952, 20140212.2273650) http://www.promedmail.org/direct.php?id=2262952 http://www.promedmail.org/direct.php?id=2273650 |
| French Guiana | 18/2/14 | Local Health authorities. http://www.invs.sante.fr/fr/content/download/84707/310791/version/54/file/pe_chikungunya_antilles_200214.pdf Promed mailing list (20140220.2290880) http://www.promedmail.org/direct.php?id=2290880 |
| Saint Kitts & Nevis | 20/2/14 | Promed mailing list (20140220.2290880) http://www.promedmail.org/direct.php?id=2290880 |
| Dominican Republic | 25/3/2014 | Promed mailing list (20140420. 2371161) http://www.promedmail.org/direct.php?id=2371161 |
| Saint-Lucia | 13/5/14 | Promed Mailing List (20140513.2467601) http://www.promedmail.org/direct.php?id=2384972 Note earlier reports of suspect cases without confirmation in January 2014 (20140129.2240929, 20140316.2335886) http://www.promedmail.org/direct.php?id=2240929; http://www.promedmail.org/direct.php?id=2335886 |
| Cuba | 11/6/14 | Promed maling list(20140614.2539532) http://www.promedmail.org/direct.php?id=2539532 |
| US Virgin Islands | 11/6/14 | Promed maling list(20140614.2539532) http://www.promedmail.org/direct.php?id=2539532 |

**Territories with confirmed importation but no reported local transmission**

| | | |
|---|---|---|
| Aruba | 24/1/14 | Promed Mailing List (20140205.2257138) http://www.promedmail.org/direct.php?id=2257138 |

**Territories with suspect –not confirmed – cases**

| | | |
|---|---|---|
| Mexico | 28/2/14 | Promed mailing list (20140302.2309812) http://www.promedmail.org/direct.php?id=2309812 |

# 4 Model for an imperfectly observed epidemic in the Caribbean

In practice, invasion statuses are imperfectly observed. Times of invasion are unobserved and there is a delay between invasion and detection. We develop here a data augmentation strategy[3-5] to tackle these problems.

Denote $d_i$ the (observed time) of the first report indicating that territory $i$ is affected. Let's assume that delay between invasion and reporting is Exponentially distributed with a mean delay of $m$ days

$$d_j - t_j \sim Exp(m)$$

## Likelihood of complete data

If the invasion status $\{I_i\}_i$, the times of invasion $\{t_i\}_i$ and of reporting $\{d_i\}_i$ were available, the likelihood of such complete dataset would be:

$$P(\{I_j, t_j, d_j\}_j | \theta) = P(\{I_j, t_j\}_j | \theta) \cdot \prod_{j: I_j=1} P(d_j | I_j, t_j, \theta) \quad (2)$$

The first term on the right hand side of equation (2) corresponds to the likelihood of the invasion process, which is given in equation (1).

The second term corresponds to the likelihood of the reporting process. For a territory that has been invaded, it is equal to

$$\prod_{j: I_j=1} P(d_j | I_j=1, t_j, \theta) = \begin{cases} \frac{1}{m} \exp\left(-\frac{(d_j - t_j)}{m}\right) & \text{if territory } j \text{ was reported as being affected} \\ \exp\left(-\frac{(T - t_j)}{m}\right) & \text{otherwise} \end{cases}$$

## Inference in a context of incomplete data

In practice, the exact time of invasion $t_j$ is unobserved and the invasion status $I_j$ is imperfectly observed. We have the following relationships:

- For a territory $j$ that has been reported as being affected, the invasion status is known ($I_j$=1). Besides, the time $t_j$ of invasion satisfies the constraint: $t_j < d_j$.
- For a territory $j$ that has not been reported as being affected, we could be in one of the following situations:
  - The territory was not invaded: $I_j$=0, $t_j$=T.
  - The territory has already been invaded, but has not been reported yet: $I_j$=1, $t_j < T$, $d_j \geq T$.

To tackle the problem of inference in this context of missing data, we implement a standard Bayesian data augmentation strategy [6]. Observed data consisting of the times of reporting $\{d_i\}_i$ are augmented with the (imperfectly observed) invasion status $\{I_i\}_i$ and the (unobserved) times of invasion $\{t_i\}_i$. The likelihood of the augmented data is given in equation (2).

We then develop a Reversible Jump Markov chain Monte Carlo strategy to explore the joint posterior distribution of parameters and augmented data [3, 4, 7]. The algorithm allows the following steps:

- Update of the parameters with a standard Metropolis-Hastings step;
- For territories that were invaded ($I_j$=1), update of the time of invasion with a standard Metropolis-Hastings step;
- For territories that were not reported as being affected, Reversible Jump steps were developed to jump between one of these two options:
    - The territory was not invaded ($I_j$=0);
    - The territory was invaded ($I_j$=1, $t_j$<$T$) but was not reported.

### Measure of goodness of fit

We measured goodness of fit by assessing how well the models agreed with the set of territories officially affected up to the time of analysis. To that end, for each model $k$, we simulated 10,000 epidemics up to the time of analysis under the sole assumption that the epidemic was seeded in Saint-Martin. We then computed the predicted probability $p_i^{(k)}$ of territory $i$ being officially affected up to the time of analysis. Our measure of goodness of fit was the log-likelihood for the set of territories officially affected

$$LL_{TOA}^{(k)} = \sum_i I_i . \log\left(p_i^{(k)}\right) + (1-I_i) . \log\left(1 - p_i^{(k)}\right)$$

where $I_i$=1 if the territory was officially affected; $I_i$=0 otherwise. The larger $LL_{TOA}$, the better the fit.

## 5  Results and sensitivity analysis

### Model comparison and impact of reporting delays

In our baseline scenario, we assume that the mean reporting delay is equal to 30 days. In a sensitivity analysis, we also consider a mean reporting delay of 0 and 15 days. We were confronted with convergence issues in the MCMC when the mean reporting delay was assumed to be 60 days.

Table S3 shows the measure of goodness of fit $LL_{TOA}$ for the different models and the different reporting delays. Overall, the fit does not appear to be much affected by assumptions about delay and the ordering of models remains unchanged.

Figure S4 presents the monthly probability of transmission as a function of distance between territories and for different reporting delays. The spatial kernel remains robust to a change in the reporting delay with transmission estimated to be only slightly more localized when the assumed reporting delay increases. Increasing the mean reporting delay increases the chance that a territory is already invaded or will be invaded soon but does not change the relative ordering of territories by risk of invasion (Figures S5).

**Table S3: Summary of the models used to study the determinants of regional spread. The covariates being considered to model the risk of transmission from territory *i* to territory *j* include: passenger flows ($T_{ij}$) or distance ($D_{ij}$) between the territories, population size of the origin territory ($P_i$) or of the destination territory ($P_j$). We also provide a measure of goodness of fit ($LL_{TOA}$) that indicates how well the model can explain the set of territories that have been officially affected as of 15 June 2014 (see section 4.3). The larger $LL_{TOA}$, the better the fit. The table presents $LL_{TOA}$ for different values of the assumed average reporting delay (0, 15 and 30 days). Our baseline scenario corresponds to a reporting delay of 30 days.**

| Model name | Instantaneous risk of transmission $\lambda_{i \rightarrow j}$ from territory *i* to territory *j* | $LL_{TOA}$ | | |
|---|---|---|---|---|
| | | **0 days** | **15 days** | **30 days (baseline)** |
| Model 0 (null) | $\lambda_{i \rightarrow j} = \beta$ | -26,52 | -26,9 | -27,28 |
| Model 1 (origin pop.) | $\lambda_{i \rightarrow j} = \beta.(P_i)^\chi$ | -26,15 | -25,9 | -25,77 |
| Model 2 (destination pop.) | $\lambda_{i \rightarrow j} = \beta.(P_j)^\chi$ | -26,92 | -27,44 | -27,82 |
| Model 3 (air transportation) | $\lambda_{i \rightarrow j} = \beta.(T_{ij})^\gamma$ | -41,08 | -40,09 | -38,7 |
| **Model 4 (distance)** | $\lambda_{i \rightarrow j} = \beta.(D_{ij})^\delta$ | **-16,98** | **-16,79** | **-16,07** |
| Model 5 (distance + origin pop.) | $\lambda_{i \rightarrow j} = \beta.(D_{ij})^\delta.(P_i)^\chi$ | -18,13 | -17,69 | -16,79 |
| Model 6 (distance + destination pop.) | $\lambda_{i \rightarrow j} = \beta.(D_{ij})^\delta.(P_j)^\chi$ | -17,03 | -16,57 | -15,83 |
| Model 7 (gravity: distance + origin pop. + destination pop.) | $\lambda_{i \rightarrow j} = \beta.(D_{ij})^\delta.(P_i)^\chi.(P_j)^\chi$ | -18,56 | -18 | -16,66 |

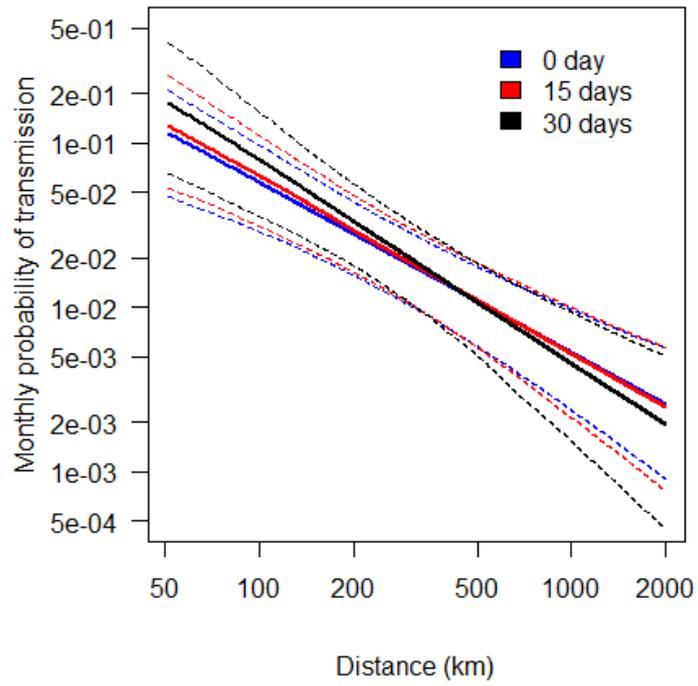

**Figure S4: Monthly probability of transmission as a function of distance between territories and for different reporting delays.**

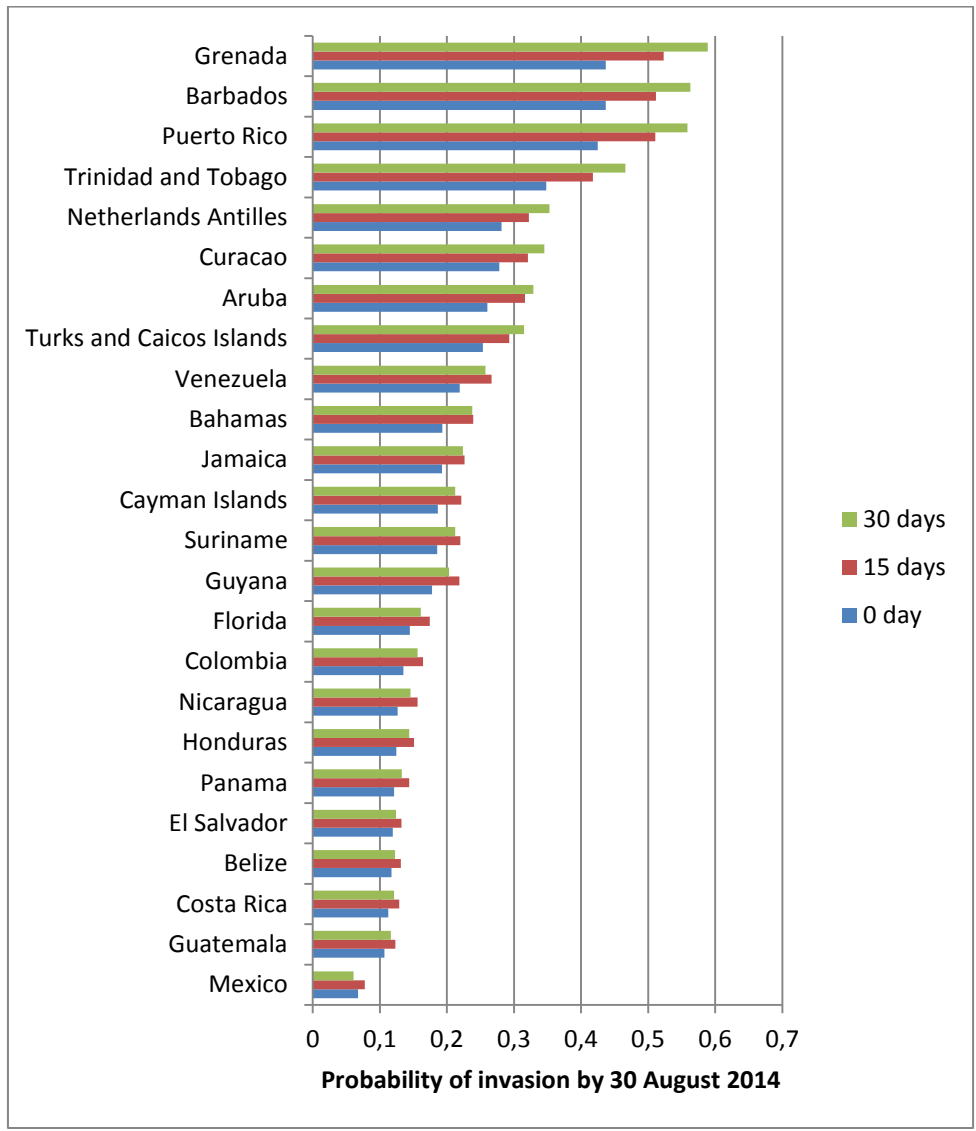

**Figure S5: Predicted probability of invasion by 30 August 2014 with data available as of 15 June 2014, as a function of the mean reporting delay. In our baseline scenario, the reporting delay is assumed to be 30 days.**

## Estimates over time

Table S4 presents the fits of the null model, the distance model and the air transportation model at different times of the epidemic. It shows that throughout the epidemic, the distance model was preferred to both the null and the air transportation model.

Table S4: Measures of goodness of fit ($LL_{TOA}$) for the null, the distance and the air transportation model with data available at different times in the epidemic. The larger $LL_{TOA}$, the better the fit. The table presents $LL_{TOA}$ for different values of the assumed average reporting delay (0, 15 and 30 days). Our baseline scenario corresponds to a reporting delay of 30 days.

|  | 0 day reporting delay | | | 15 day reporting delay | | | 30 day reporting delay | | |
| --- | --- | --- | --- | --- | --- | --- | --- | --- | --- |
|  | Null | Distance | Air transp. | Null | Distance | Air transp. | Null | Distance | Air transp. |
| 15/01/2014 | -18,64 | -13,16 | -22,68 | -17,66 | -11,67 | -20,86 | -15,85 | -10,02 | -21,56 |
| 15/02/2014 | -18,2 | -10,85 | -33,48 | -17,66 | -10,13 | -34,77 | -17,67 | -9,15 | -30,84 |
| 15/03/2014 | -20,11 | -13,07 | -49,2 | -19,89 | -12,7 | -48,93 | -19,82 | -12,03 | -43,71 |
| 15/04/2014 | -21,08 | -14,59 | -41,88 | -21,17 | -14,11 | -41,44 | -21,29 | -13,83 | -37,25 |
| 15/05/2014 | -24,8 | -16,42 | -43,66 | -24,9 | -15,96 | -43,12 | -25,23 | -15,38 | -39,96 |
| 15/06/2014 | -26,52 | -16,98 | -41,08 | -26,9 | -16,79 | -40,09 | -27,28 | -16,07 | -38,7 |

Figure S6 presents the estimated transmission kernel at different time points of the epidemic. Earlier estimates favoured slightly more short range transmission when compared with later estimates.

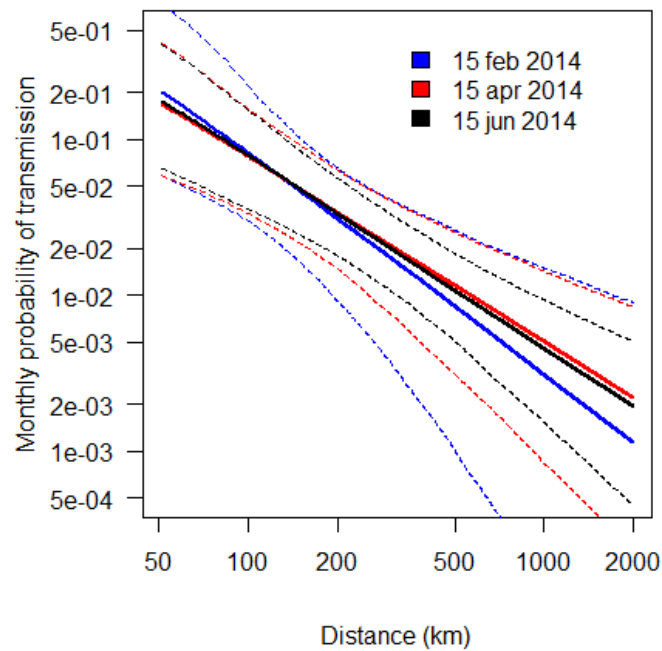

Figure S6: Monthly probability of transmission as a function of distance between territories estimated at different times in the epidemic.